\documentclass{arxiv}
\usepackage{graphicx}
\ifCUPmtlplainloaded
\else
  \font\tenbms=cmbsy10          \skewchar\tenbms ='60
  \font\sevenbms=cmbsy10 at 7pt \skewchar\sevenbms ='60
  \font\fivebms=cmbsy10 at 5pt  \skewchar\fivebms ='60
  \newfam\bmsfam
  \textfont\bmsfam=\tenbms
  \scriptfont\bmsfam=\sevenbms
  \scriptscriptfont\bmsfam=\fivebms
  
\newcommand{\etal}{{\it et al\/}.\ }
\title{On coherent structures in the compressible turbulent round jet}
\author[P. Reynier and H. Ha Minh]{P.\ns  R\ls E\ls Y\ls N\ls I\ls E\ls R\ns 
\and H.\ns  H\ls A\ns M\ls I\ls N\ls H}

\affiliation{Institut de M\'ecanique des Fluides de Toulouse, \\
       Avenue du Pr. Camille Soula, 31400 Toulouse, France}
\pubyear{1997}
\volume{0}
\date{}
\begin{document}
\maketitle

\tolerance 10000

\begin{abstract}
The presence of two-dimensional coherent structures in the near-field of 
the round jet has been established by several experimental investigations 
and direct Navier-Stokes simulations (DNS). Their study has a great importance 
to improve the prediction of unsteady flows dominated by large scale structures. 
Indeed, coherent structures play a 
determinant role on turbulence and flow evolution. The goal of this numerical 
study is to  apply the 
semi-deterministic modelling, a method close to the large eddy simulation 
(LES), to the simulation of compressible round jets. This allows the 
simulation of natural unsteadiness without any flow excitation. A free jet 
configuration is computed for three Mach numbers of 0.3, 0.96 and 1.5 to
evaluate the influence of compressibility effects on instability, Strouhal 
number and expansion rate. There is not quantitative comparisons on 
flow quantities with experimental data, except for the Strouhal number, due to 
the discrepancies between the configuration computed here and those experimentally 
investigated in past studies. The numerical results on coherent structures 
are validated across a comparison on the Strouhal number of the preferred mode. 
The results on the evolution of the Strouhal number with the Mach number are in 
excellent agreement with previous experiments. Moreover, the evolution of the large 
scale vortices in the flow is well correlated by experimental and DNS results 
obtained on other configurations.
\end{abstract}

\section{Introduction}

Since the pioneering works of \cite{LR879} the jet instability has been the 
matter of many experimental and theoretical 
studies (e.g. Crow \& Champagne 1971; Lau \& Fisher 1975; 
Hussain \& Zaman 1981; Sokolov, Kleis \& Hussain 1981; 
Liepmann \& Gharib 1992). All these studies turned on the investigation on 
large scale structures in the round jets. The coherent structures such as they 
have been defined by Hussain (1983, 1986) exist in the round jet as in any flow 
including a mixing layer. Their discovery in the near-field of the axisymmetric 
jet is the fact of \cite{CC71}. 
Their works put in evidence a preferred mode for this flow with 
an associated Strouhal number of about $0.3$. Later, \cite{HZ81} investigated more 
largely this preferred mode and established its independence 
with the initial boundary layer instability. Indeed, the 
characteristic length of this mode is not the boundary layer 
thickness at the inlet pipe but the jet diameter. Large scale vortices evolve 
in the jet shear-layer and \cite{LF75} showed that the basic 
structure of the turbulent round jet consists essentially in an alley of 
vortices moving downstream in the mixing layer.

In the round jet, the entrainment process is drastically influenced and increased by the 
large scale structures evolving in the shear-layer (see Liepmann 
\& Gharib 1992). In this way, if the entrainment is constant in 
the far-field, it increases in the near-field. \cite{RS61} suggested a linear 
growth for x/D$\leq$5 which agrees with the experimental 
results of \cite{LG92}. For \cite{BP79}, the entrainment rate is constant in 
the near region. This discrepancy between these experimental results can be 
explained by different inlet conditions: in the experiment of 
\cite{BP79} the flow is fully-developed whereas in that of \cite{LG92} 
the turbulence level at the exit is lower (young turbulence). Moreover, these  
last authors established that the entrainment capacity of a laminar or transitional 
jet is larger than that of a fully-developed turbulent jet. This is a consequence 
of the presence  of coherent 
structures which carry some fluid across the jet interface involving a large growth 
of the shear-layer. On the other hand, these vortices increase the interface 
length, therefore the diffusion between the jet and the outer environment. 

If the greater part of experiments on jet instabilities 
have been carried out for low or moderate Reynolds numbers and incompressible 
flows (Crow \& Champagne 1971; Lau \& Fisher 1975; Hussain \& Zaman 1981), 
some studies have been led for high Reynolds 
number or compressible flows (e.g. Arnette, Samimy \& Elliot 1993;  
Fourguette, Mungal \& Dibble 1991; Lepicovski \etal 1987). 
They established the existence of coherent 
structures even for high Mach numbers: $2.5$ in the experiment of \cite {MM80}. 
\cite{MO77} had already demonstrated that the large scale structures 
present at low Reynolds numbers $(10^{4})$ are also observed at Reynolds numbers 
of about $10^{6}$. These coherent structures, present in the mixing layer of the jet, 
play a central role in the flow, revealed by the fundamental works of \cite{BR74} and \cite{WB74} 
on the compressible mixing layer. The two-dimensional coherent 
structures and their relation to the hydrodynamic instability have been 
studied for the following reasons:

\begin{itemize}
\item[-] Their role in the laminar-turbulent transition;
\item[-] The interest for the theoretical modelling to improve the prediction 
of flow and mixing.
\item[]
\end{itemize}

The vortices evolving in the shear-layer originate from instability waves 
of which the growth is ensured by the nonlinear interactions of the flow. This 
phenomenon has been experimentally investigated by \cite {LG92} and numerically by 
\cite {VO94} and Grinstein, Oran \& Hussain (1987) for incompressible and 
compressible jets.
There are few studies on instability waves in compressible jets. Indeed, 
the measurements on pressure and velocity fluctuations, more 
particularly in  supersonic flows, do not give impressive results which 
explains the restricted number of experiments carried out. However, \cite{OER82} 
and \cite{TH89} observed that three different families 
of instability waves, with distinct wave patterns and propagation characteristics, 
are present in supersonic jets:

\begin{itemize}
\item[-] The Kelvin-Helmhotz instability;
\item[-] The subsonic waves (waves with subsonic phase velocities);
\item[-] The supersonic waves (waves with supersonic phase velocities).
\item[]
\end{itemize}

These three modes of instability modes are convective. \cite{TH89} 
did not observe absolute instabilities in their investigations. When the 
jet is subsonic, only the Kelvin-Helmotz instability is active. This phenomenon 
is independent from the Reynolds number: \cite{LUBB87} 
showed that Kelvin-Helmotz instability waves are present at Reynolds numbers 
higher than $10^{6}$. On the other hand, the growth rate of this 
instability kind decreases when the Mach number increases (Miles 1958). 
The subsonic waves are present for low supersonic jets, the last mode of 
instability, the supersonic waves, is present if only the Mach number is higher than 
the sum of ambient and jet speeds of sound (Tam \& Hu 1989).

In the present paper only the Kelvin-Helmoltz instability is simulated. The involving 
waves are at the origin of the coherent structures which control the 
dynamic and the mixing of the jet. According to \cite{TH89} the flow is 
dominated by a similar process for low and moderate supersonic Mach numbers. 
\cite{ZL92} showed that the 
large scale vortices originate from pressure fluctuations at the jet interface. The 
results of \cite{MI84}, \cite{LG92} and \cite{VO94} show that the shear-layer 
becomes unstable near the inlet then rolls up to form vortex rings and finally turns 
fully turbulent.

The mixing layer vortices appear before x/D$=$2 (e.g. Arnette \etal 1993; Grinstein 
\etal 1987; Liepmann \& Gharib 1992), next both 
contraction and expansion enhance the evolution 
of three-dimensional secondary structures. The experimental 
investigation of \cite{LG92} as well as the DNS results of \cite{VO94} show 
that three-dimensional structures in finger-form 
appear on the edges of the two-dimensional vortices with fluid ejection in the shape 
of lateral jets. These three-dimensional effects are strong downstream the potential 
core at 4D$\leq$x$\leq$5D (e.g. Moore 1977; Sokolov \etal 1981; 
Grinstein \etal 1987; Liepmann \& Gharib 1992). They 
result from the strongly nonlinear interactions of the flow-field. In 
supersonic jets, the three-dimensional unsteadiness have been observed at the 
end of the potential core by \cite{MM80}. For \cite{FMD91} the supersonic jets 
are more three-dimensional than incompressible jets, due to oblique 
instability waves in the flow. \cite{CM92} and Samimy, Reeder \& Elliot (1992) 
have also noted the same phenomeon in the compressible mixing 
layer. The stability analysis of \cite{RW89} shows that the 
oblique instability modes are more amplified than the two-dimensional modes 
in highly compressible flows.

The aim of this paper is essentially to improve the prediction of the jet 
near-field, especially taking into account the two-dimensional 
coherent structures. The three-dimensional aspects will not be takle here. 
For the turbulence modelling, since DNS and LES are not available for the 
numerical computation of high Reynolds jets,  the semi-deterministic 
approach developed by \cite{HK93} and already applied to the  
simulation of coaxial jets by \cite{RH95} has been choosen. This 
method is close to the LES but differs slightly by its spirit. The main 
purpose of this work is to apply the semi-deterministic modelling (SDM) 
to simulate natural  flow 
unsteadiness in a compressible round jet. Therefore, main features on 
coherent structures as their evolution in the flow and their interaction 
with the mean quantities are studied. Moreover,  the influence of compressibility 
effects on coherent structures and Strouhal number is evaluated. 
So, several 
calculations have been led on the same configuration for several Mach numbers, 
they allowed the prediction of dynamic and energetic quantities in the 
computational flow. Finally, to verify the trends observed, these numerical 
results are qualitatively compared to those of 
experimental and DNS investigations.

\section{Methodology}

\subsection{The semi-deterministic modelling}

Most of flows contain organized and random (in the sense of chaotic) 
characters. For a long time, the numerical predictions have been limited to 
the calculation of approached problems (Prandtl approximation) where the 
flow unsteadiness is not considered. So, the spreading out of the 
turbulence closures in the seventies has been built on the resolution 
under these approximations. Therefore, turbulence statistical models lead to 
satisfying results when the turbulent energy is in spectral equilibrium, the 
turbulence is then fully-developed. However, many 
flows particularly in industrial problems are rather concerned by young 
turbulence (i.e. near the transition). This last is 
characterized by two main features:

\begin{itemize}
\item[-] The turbulence may be at low  turbulent Reynolds numbers so the turbulent 
 and molecular diffusions may be of the same order;
\item[-] The three-dimensional spectrum of the energy is not in 
equilibrium, it is strongly modified by the presence of unsteady organized 
structures.
\item[]
\end{itemize}

These organized structures constitute the coherent part of the turbulence. 
They are highly dependent on initial and boundary conditions and they 
strongly influence the turbulence energetic level. Indeed, \cite{HZ81}
and \cite{SKH81} remarked in their experimental 
investigations on round jets (with inlet conditions characterized by low 
turbulence levels) that near the exit the 
coherent part of the turbulent energy is greater than the random 
component. Therefore, the classical statistical modelling need to be reconsidered 
to take account of the organized unsteadiness. 

Reynolds \& Hussain (1972) demonstrated that every instantaneous physical quantity 
$f(x_{k},t)$ can be split in three components:
\begin{equation}
f(x_{k},t) = \underbrace{\overline{f}(x_{k})}_{(a)} + 
\underbrace{{f_{c}}(x_{k},t)}_{(b)} 
+ \underbrace{{f_{r}}(x_{k},t)}_{(c)}
\end{equation}

The part (a) represents the time-averaged quantity. It is generally the only 
quantity available from experiment. The coherent or organized 
unsteadiness (b) posesses a determinist character available from numerical 
calculations. Coherent unsteadiness is highly dependent on the fundamental 
properties of the flow: geometry, boundary and initial conditions 
and physical properties of the fluid medium. The component (c) points 
out the random or incoherent fluctuations. The corresponding structures 
are characterized by a continuous spectrum which some peaks, traces of 
organized structures may be surimposed. In theory, the simulation of 
all vortices (DNS) even for small scales is possible. But, if DNS is the most 
natural way to predict a flow, the computation cost of this pure approach is too 
expensive for high Reynolds number flows. Finally, the parts whose the computation 
can be reasonably 
expected are the time-averaged motion and the coherent unsteadiness. 
A new quantity, the phase average, is obtained. Determined from the successive coherent 
structures at the same age (or phase) of their evolution, its expression is:
\begin{equation}
\langle f(x_{k},t)\rangle = \overline{f}(x_{k}) + {f_{c}}(x_{k},t) 
\end{equation}
so, 
\begin{equation}
f(x_{k},t) = \langle f(x_{k},t)\rangle + {f_{r}}(x_{k},t) 
\end{equation}

If $f_{r}=0$, the DNS is found. The case where $f_{r}$ is 
very small corresponds to the LES approach. Finally, if the quantity 
$\langle f(x_{k},t) \rangle$ is minimized and reduced to the time average,  
the classical statistical approach is obtained. Also, the SDM appears as an intermediate way 
between the classical statistical modelling and the DNS. Although this 
approach is very close to the LES, several differences exist between these two 
methods. From the methodology point of view some questions on the necessity to 
distinguish LES and SDM can be legitimately set. In the formalism field, 
the equations to solve for the two methods are the same. Indeed, the phase average 
leads to the same equations that the classical modelling. The only differences 
in the governing equations come from the closure, sub-grid modelling  for 
the LES while $f_{r}$ is represented by a turbulence model in SDM. The main discrepancy 
between the two approaches is the splitting used for the vortical structures. 
The LES splits the structures according to their size: large scale structures to 
be simulated and small scale structures to be modelled. SDM splits the vortices in 
function of their nature: coherent structures to be computed and random 
turbulence to be modelled. If the coherent structures have a highly 
two-dimensional character, a two-dimensional approach  much economical can be used with 
SDM, while LES must use three-dimensional calculations. As the part to model 
contains all the random 
components the situation is nearly identical to the classical modelling. 
The governing equations are similar, so the use of the same process to 
built the new closures is tempting. However,  some modifications must be 
done in the set of constants. According to \cite{HK93}, as the turbulence role has 
been defined again, the constants must be recalibrated.

\subsection{Governing equations}

The near-field of the round jet is highly influenced by the presence of 
two-dimensional large coherent structures if the initial conditions allow 
it (under-developed turbulence or laminar flow), so the present study is limited 
to the two-dimensional aspect of this flow. The governing equations are the 
unsteady Navier-Stokes equations written in axisymmetric coordinates with the 
mass-weighted average of Favre (1965) and a state equation for a perfect gas. 
The closure is obtained using a 
$k-\epsilon$ turbulence model. The governing equations for the conservation of 
density $\overline{\rho}$, pressure $\overline{P}$, momentum 
$\overline{\rho}\tilde{U}$ in streamwise 
direction x, momentum $\overline{\rho}\tilde{V}$ in spanwise direction r 
and total energy $\tilde{E}$ (the equations of the turbulence model 
will be presented in \S\,2.3) are\\

{\it Equation for the conservation of the mass-density $\overline{\rho}$} 
\begin{equation}
{\rm \partial  \over \partial t}\overline{\rho }+{\partial \over \partial 
{x}}\overline{\rho }{\tilde{U}}+{1 \over r}{\partial  \over \partial r}r
\overline{\rho }{\tilde{V}}=0
\end{equation}

{\it Equation for the conservation of the momentum $\overline{\rho}\tilde{U}$}  
\begin{equation}
\matrix {\displaystyle
{\rm \partial  \over \partial t}\overline{\rho }
{\tilde{U}}+{\partial  \over \partial {x}} \biggl\{ \overline{\rho }
{\tilde{U}}{\tilde{U}}+{2 \over 3}\overline{\rho }
\tilde{k}+\overline{P}-(\mu +{\mu }_{t}){\tilde{S}}_{xx} \biggr\} \cr \cr 
\displaystyle
+{1 \over r}{\partial  \over \partial {r}}r \biggl\{ \overline{\rho 
}{\tilde{U}}{\tilde{V}}-(\mu +{\mu }_{t}){\tilde{S}}_{rx} \biggr\} =0\cr}
\end{equation}

{\it Equation for the conservation of the momentum $\overline{\rho}\tilde{V}$}  
\begin{equation}
\matrix{\displaystyle{ \rm \partial  \over \partial t}\overline{\rho 
}{\tilde{V}}+{\partial  \over \partial {x}} \biggl\{ \overline{\rho 
}{\tilde{U}}{\tilde{V}} -(\mu +{\mu }_{t}){\tilde{S}}_{xr} \biggr\} \cr \cr \displaystyle
+{1 \over r}{\partial  \over \partial {r}}r
 \biggl\{ \overline{\rho }{\tilde{V}}{\tilde{V}}+\overline{P}
+{2 \over 3}\overline{\rho }\tilde{k}-(\mu +{\mu 
}_{t}){\tilde{S}}_{rr} \biggr\} =\cr \cr \displaystyle
{1 \over r} \biggl\{ \tilde{P}+{2 \over 3}\overline{\rho }\tilde{k}-(\mu +{\mu 
}_{t}){\tilde{S}}_{\theta \theta } \biggr\} \cr}
\end{equation}

\noindent
The terms of the strain tensor are given by the following expressions:

\begin{equation}
{\tilde{\rm S}}_{rr}=2{\partial \tilde{V} \over \partial r}-{2 \over 
3} \left({\partial \tilde{U} \over \partial x}+{\partial \tilde{V} \over 
\partial r}+{\tilde{V} \over r}\right)
\end{equation}

\begin{equation}
{\tilde{\rm S}}_{xx}=2{\partial \tilde{U} \over \partial x}-{2 \over 
3} \left({\partial \tilde{U} \over \partial x}+{\partial \tilde{V} \over 
\partial r}+{\tilde{V} \over r}\right)
\end{equation}

\begin{equation}
{\tilde{\rm S}}_{xr}={\tilde{S}}_{rx}={\partial \tilde{U} \over \partial 
r}+{\partial \tilde{V} \over \partial x}
\end{equation}

\begin{equation}
{\tilde{\rm S}}_{\theta \theta }=2{\tilde{V} \over r}-{2 \over 
3}\left({\partial \tilde{U} \over \partial x}+{\partial \tilde{V} \over 
\partial r}+{\tilde{V} \over r}\right)
\end{equation}

{\it Equation for the conservation of the total energy $\tilde{E}$} 
\begin{equation}
\matrix{\displaystyle{\rm \partial  \over \partial t}\overline{\rho }\tilde{E}+
{\partial  \over \partial x}\biggl\{\biggl(\overline{\rho }\tilde{E}+\overline{P}
+{2 \over 3}\overline{\rho }\tilde{k}
-(\mu +{\mu }_{t}){\tilde{S}}_{xx}\biggr)\tilde{U}\cr \cr \displaystyle
-(\mu +{\mu }_{t}){\tilde{S}}_{xr}\tilde{V}
-\gamma \biggl({\mu  \over Pr}+{{\mu }_{t} \over {Pr}_{t}}\biggr){{\partial } \over
 \partial x}\tilde{{e}_{i}} \biggr\}\cr \cr \displaystyle
+{1 \over r}{\partial  \over \partial r}r\biggl\{ \biggl(\overline{\rho }\tilde{E}
+\overline{P}+{2 \over 3}\overline{\rho }\tilde{k}
-(\mu +{\mu }_{t}){\tilde{S}}_{rr} \biggr)\tilde{V}\cr \cr \displaystyle
-(\mu +{\mu }_{t}){\tilde{S}}_{xr}\tilde{U}
-\gamma \biggl({\mu  \over Pr}+{{\mu }_{t} \over {Pr}_{t}}\biggr){\partial\over
 \partial r}\tilde{{e}_{i}} \biggr\}=0\cr}
\end{equation}

{\it State equation for a perfect gas}
\begin{equation}
\tilde{\rm P}=(\gamma -1)\overline{\rho }\tilde{{e}_{i}}
\end{equation}

Where $\mu$ is the molecular viscosity, $\mu _{t}$ the turbulent viscosity, 
$Pr$ the Prandtl number, $Pr _{t}$ the turbulent Prandtl 
number, $\tilde{e_{i}}$ the internal 
energy and $\gamma$ the specific heat ratio.

\subsection{Turbulence model}

A $k-\epsilon$ model is used for the turbulence modelling. Generally, this 
level of 
closure gives good predictions for thin shear flows. The usual set of constants 
(Launder \& Sharma 1974), "supposed 
universal", is not calibrated for flows where organized structures are 
present but for fully-developed turbulence. So, to take care of coherent 
unsteadiness a new calibration is necessary. The most debated constant is 
$C_{\mu}$ whose the usual value of $0.09$ leans on the hypothesis of equilibrium 
between the sources ($Production = Dissipation$). The experimental investigation 
of \cite{RO72} on a round 
jet has shown the failure of this assumption in the near-field of this 
flow. This is a consequence of the organized unsteadiness which influences 
the energetic balance in this region. In fact, in flows where organized 
structures evolve, the normal stresses are 
larger and the turbulent shearing is weaker than in flows in equilibrium. 
In consequence, the basic value of $C_{\mu}$  is too high. \cite{RO72} had 
already remarked the inconstant character of $C_{\mu}$ and he 
had exprimed this quantity in function of its usual value and of the ratio 
between the production and the 
dissipation. Later, \cite{PO75} proposed an effective viscosity approach 
for the $k-\epsilon$ 
model and a nonlinear function for $C_{\mu}$ to include some anisotropic effects. 
But, with these two propositions the  model is then numerically more unstable. 
Therefore, in the present paper the value of $C_{\mu}$ proposed by \cite{HK93} 
is retained. They have recalibrated this constant on a backward-facing step 
at a value of $0.02$. The main advantage of this lower value is to imply a 
weaker numerical dissipation due to the lower diffusion of the model, 
therefore the prediction of the vortical 
structures is easier. A second  advantage of this choice is to reduce the production of the 
turbulent kinetic energy generally overestimated by the $k-\epsilon$ 
model. There is no additive terms modelling compressibility effects in this study. 
According to \cite{SEHK91} the incompressible turbulence models are expected to 
give good results for the jets up to a Mach number of 1.5. The equations for turbulent 
kinetic energy and the dissipation rate can be written under the following form:\\

{\it Equation of turbulent kinetic energy}
\begin{equation}
{\rm \partial  \over \partial t}\overline{\rho }\tilde{k}+{\partial  
\over \partial x}\biggl\{\overline{\rho }\tilde{k}\tilde{U}-{\mu }_{k}{\partial  
\over \partial x}\tilde{k} \biggr\}
+{1 \over r}{\partial  \over \partial 
r}r\biggl\{\overline{\rho }\tilde{k}\tilde{V}-{\mu }_{k}{\partial  \over \partial 
r}\tilde{k} \biggr\}=
{\tilde{P}}_{k}-\overline{\rho }\tilde{\varepsilon}+W_{k}
\end{equation}
with   
\begin{equation}
{\mu}_{k} = {\mu}+{{\mu}_{t}\over{\sigma}_{k}}
\end{equation}

\begin{equation}
W_{k} = -2{\mu}{\mu}_{t} \left ({\partial ({\tilde{k})^{1 \over 2}}\over \partial 
x_{n}} \right )^{2}
\end{equation}\\

\noindent
The production term expression is

\begin{equation}
\matrix{\displaystyle{\tilde{\rm P}}_{k}=2{\mu }_{t}\biggl\{\biggl({\partial 
\tilde{U} \over \partial x}{\biggr)}^{2}+\biggl({\partial \tilde{V} \over 
\partial r}{\biggr)}^{2}+\biggl({\tilde{V} \over r}{\biggr)}^{2}\biggr\}
+{\mu }_{t}\biggl({\partial \tilde{U} \over \partial r}+{\partial \tilde{V} 
\over \partial x}{\biggr)}^{2}\cr \cr \displaystyle
-{2 \over 3}\biggl({\partial \tilde{U}\over \partial x}+{\partial \tilde{V} 
\over \partial r}+{\tilde{V} \over r}\biggr) 
\biggl\{ {\mu }_{t}\biggl({\partial \tilde{U} \over \partial x}+{\partial \tilde{V} 
\over \partial r}+{\tilde{V} \over r}\biggr)+\overline{\rho }\tilde{k}\biggr\}\cr}
\end{equation}\\

{\it Equation for the dissipation rate of the turbulent kinetic energy}
\begin{equation}
\matrix{\displaystyle{\rm \partial  \over \partial t}\overline{\rho 
}\tilde{\varepsilon }+{\partial  \over \partial x}\biggl\{\overline{\rho 
}\tilde{\varepsilon }\tilde{U}-{\mu }_{\varepsilon }{\partial  \over 
\partial x}\tilde{\varepsilon }\biggr\}
+{1 \over r}{\partial  \over \partial 
r}r\biggl\{\overline{\rho }\tilde{\varepsilon }\tilde{V}-{\mu }_{\varepsilon 
}{\partial  \over \partial r}\tilde{\varepsilon }\biggr\}=\cr \cr \displaystyle
{C}_{{\varepsilon }_{1}}{\tilde{P}}_{k}-{C}_{{\varepsilon 
}_{2}}{\overline{\rho }{\tilde{\varepsilon }}^{2} \over \tilde{k}}+W_
{\varepsilon}\cr}
\end{equation}
with   
\begin{equation}
{\mu}_{\varepsilon} = {\mu}+{{\mu}_{t}\over{\sigma}_{\varepsilon}}
\end{equation}

\begin{equation}
W_{\varepsilon} = {-2{\mu}{\mu}_{t}\over \overline{\rho}} \left ({\partial ^{2}
 {\tilde{V}}\over \partial { x_{n}}^{2}} \right )^{2}
\end{equation}

\noindent
Where $x_{n}$ is the distance to the wall (here x), $W_{k}$ and $W_{\epsilon}$ are only 
active in the near wall regions, therefore in the present calculations these terms 
are negligible. The set of constants of the turbulence model is

\begin{equation}
C_{\mu} = 0.02  ;  {C}_{{\varepsilon }_{1}} = 1.44 ; 
{C}_{{\varepsilon }_{2}} = 1.92 ; \sigma_{k} = 1. ; 
\sigma_{\varepsilon} = 1.3
\end{equation}

\section{The numerical method}

\subsection{The computed configuration}

An air-air configuration (represented in figure 1) is  predicted for  several 
Mach numbers: a supersonic, a transonic and a subsonic (quasi-incompressible).  
The inlet conditions are derived from the experiment of \cite{DU71} which is 
characterized by a lower turbulence level on the jet axis than in a fully-developed jet as the 
one investigated by \cite{CH79}. 
The different computed cases are resumed in the table 1.

\begin{table}
  \begin{center}
       \begin{tabular}{cccc}
Velocity $U_{0}$ (m/s) & Mach number & Reynolds number & Inlet conditions  \\[3pt] 
$104$ & $0.3$ & $52240$ & Dur\~{a}o \\
$333$ & $0.96$ & $167200$ & Dur\~{a}o \\
$520$ & $1.5$ & $261200$ & Dur\~{a}o \\
     \end{tabular}
     \caption{Computed cases}
  \end{center}
\end{table}

The injector diameter D is $7.24$ mm. The exit temperature is 
$300^{o}$K, the pressure $P_{e}$ at the exit and in the computational field is 
initially of $0.101$ MPa and the initial density in all the field is 
$\rho_{o}=1.28$ kg.m$^{-3}$. The computation used 100 $\times$ 93 grid points with  
stretching in the transverse direction outside the jet. The mesh is uniform 
in the x direction. The computational domain extends over $16.6$ diameters in the streamwise 
direction and over $8.3$ diameters in the transverse direction.

\begin{figure}
  \begin{center}
\includegraphics[width=80mm]{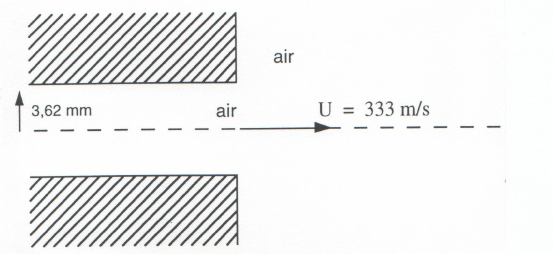}
    \caption{The axisymmetric configuration for the transonic jet}
    \end{center}
\end{figure} 

\subsection{Numerical scheme}

The numerical method used for the computations is the finite volume scheme 
proposed by \cite{MC81}. It is an explicit-implicit method, 
parabolic in time and elliptic in space. In the implicit part the 
treatment of viscous terms is approximated, so this algorithm is more suitable to 
the prediction of steady than unsteady problems. In consequence, for the 
present simulations only the explicit part is used in order to predict the flow 
instability. The equations are resolved in 
conservative form and the prediction-correction step procedure is utilized. 
This scheme is accurate to second order in time and space and does not require 
any additional numerical dissipation for its stability. This last skill is 
fundamental to the simulation of natural flow unsteadiness. Indeed, the presence 
of numerical dissipation would involve a large diffusion which 
would prevent the simulation of coherent structures. 
In this paper, there is no comparison of the numerical results with experiments 
on the mean quantities. This is due to the discrepancies between the flow pattern 
computed and those experimentally investigated in past studies. Nevertheless, 
in a previous numerical simulation (e.g. Reynier \& Ha Minh 1996), a validation of our code 
with respect to the experimental data of \cite{RI72} have been led for incompressible 
coaxial jets.

\subsection{Boundary conditions}

The inlet conditions are derived from the experimental 
data of \cite{DU71}. The dissipation rate is initialized considering the flow 
at the inlet as a boundary layer in equilibrium. Outside the
jet, a wall is present (see figure 1) in the radial direction, 
so Dirichlet boundary conditions (zero) 
are applied on this boundary for pressure and density gradients, velocity, turbulent 
kinetic energy and dissipation rate. The lower boundary (jet axis) is 
a symmetry axis so zero gradients are assumed for all quantities. 
For the upper boundary two different conditions have been used: a) Neumann conditions 
(zero gradients) for all quantities ; b) symmetry conditions (if the flow takes place 
in the context of a multi-jets configuration where one jet is surrounded by several others). 
As the upper boundary is far 
from the jet interface there is no influence of this boundary condition on the 
flow unsteadiness.  As the present study is carried 
out in the perspective of turbulent modelling for injection in rocket engines, 
the computation results presented in this paper have been obtained with 
symmetry conditions.  The outlet 
conditions are deducted from characteristic relationships. They originated from 
the characteristic analysis theory and they have been developed for the Euler 
equations by \cite{TH87}. These non-reflecting boundary conditions are:

\begin{equation}
{\rm \partial \overline{\rho } \over \partial t}-{1 \over 
{c}^{2}}{\partial \overline{P} \over \partial t}=
-\tilde{U}\left({{\partial \overline{\rho } \over \partial x}-{1 \over 
{c}^{2}}{\partial \overline{P} \over \partial x}}\right)
\end{equation}

\begin{equation}
{\rm \partial \overline{P} \over \partial t}+\overline{\rho }c{\partial 
\tilde{U} \over \partial t}=
-\left({\tilde{U}+c}\right)\left({{\partial \overline{P} \over \partial x}
+\overline{\rho }c{\partial \tilde{U} \over \partial x}}\right)
\end{equation}

\begin{equation}
{\rm \partial \tilde{V} \over \partial t}=-\tilde{U}\left({{\partial 
\tilde{V} \over \partial x}}\right)
\end{equation}

\begin{equation}
{\rm \partial \overline{P} \over \partial t}-\overline{\rho }c{\partial 
\tilde{U} \over \partial t}=
-\left({\tilde{U}-c}\right)\left({{\partial \overline{P} \over \partial x}
-\overline{\rho }c{\partial \tilde{U} \over \partial x}}\right)
\end{equation}

When the flow is subsonic the pressure must be specified at the exit, 
a zero gradient is applied for this quantity. For the turbulent 
kinetic energy and its dissipation rate the same conditions are used on this 
boundary.

\section{Results}

Section \S\,4.1 focuses on presence of instabilities and evolution 
of the coherent structures in the near-field of the jet. The evolution 
of flow quantities (velocity, pressure, density, turbulent kinetic energy and 
dissipation rate) in the flow-field and their interaction with the organized 
unsteadiness are presented in \S\,4.2. The compressibility effects and their 
influence on large scale structures are studied in section \S\,4.3. A comparison 
between the Strouhal numbers obtained experimentally and our numerical results 
is shown in this last section.
 
\subsection{The unsteady round jet}

\subsubsection{Presence of unsteadiness}

\begin{figure}
  \begin{center}
\includegraphics[width=80mm]{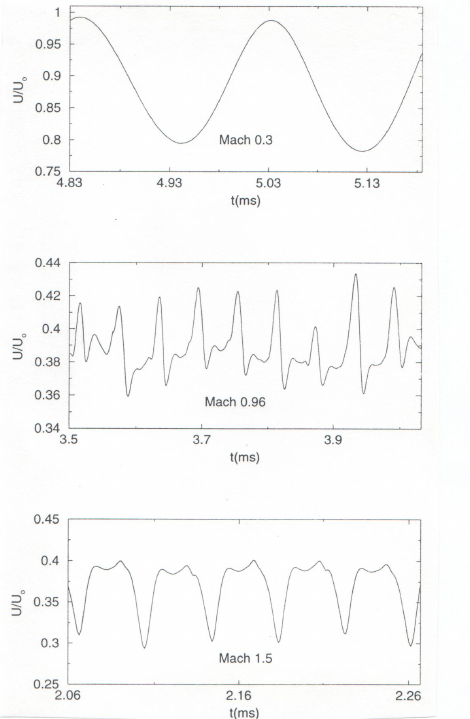}
 \caption{Time-variations of the streamwise velocity in the near-field at x$=$1.5D
 and y$=$0.5D: a) Mach 0.3; b) Mach 0.96; c) Mach 1.5}
    \end{center}
\end{figure}

\begin{figure}
  \begin{center}
\includegraphics[width=80mm]{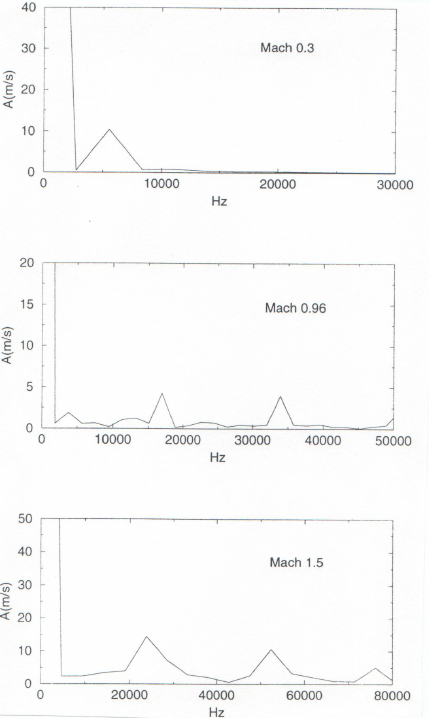}
 \caption{Spectra of time-variations of the streamwise velocity in the near-field at 
x$=$1.5D and y$=$0.5D: a) Mach 0.3; b) Mach 0.96; c) Mach 1.5}
    \end{center}
\end{figure}

The executed computations with inlet conditions derived from the experimental 
data of \cite{DU71} lead to the simulation of instabilities (see figure 2) in the 
near-field without any flow excitation. The figure 2 represents the time-dependent 
variations of the streamwise velocity for a point located in the 
shear-layer at x$=$1.5D and y$=$0.5D. This quasi-periodic phenomenon is independent 
from the Mach number. If for a Mach number of 0.3 the variations are 
quasi-sinusoidal, this is not the case for the transonic and supersonic jets, 
due to the presence of a pairing near this location. The corresponding spectra 
obtained by Fourier 
analysis over one hundred periods are presented in figure 3. They show a dominant 
frequency for the unsteadiness and some harmonics for the two higher Mach numbers. 
The dominant frequency is equal to 5600 Hz for the 
quasi-incompressible case, 17500 Hz for the transonic jet  
and  23800 Hz for the supersonic jet. These instabilities correspond to the 
dominant  large scale structure evolving in the mixing layer. The associated 
Strouhal numbers (calculate from the diameter of the inlet pipe and the exit 
velocity) are 0.39 for a Mach number of 0.3, 0.38 for the transonic jet  
and  0.33 for the supersonic jet. The Strouhal numbers are in the range of 
values contained between 0.3 and 0.4 corresponding to the preferred mode 
observed in the experiments of \cite{CC71} and \cite{HZ81}, and predicted by 
the theory (e.g. Michalke 1984). 
This preferred mode corresponds to the coherent structures which dominate the 
shear-layer of a round jet.

\subsubsection{Coherent structure evolution}

\begin{figure}
  \begin{center}
\includegraphics[width=120mm]{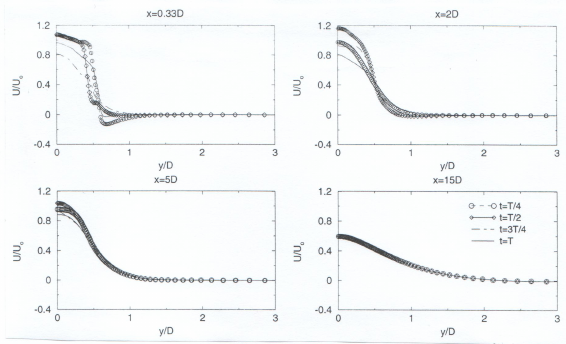}
   \caption{Profiles of the unsteady streamwise velocity for the transonic jet}
    \end{center}
\end{figure} 

\begin{figure}
  \begin{center}
\includegraphics[width=120mm]{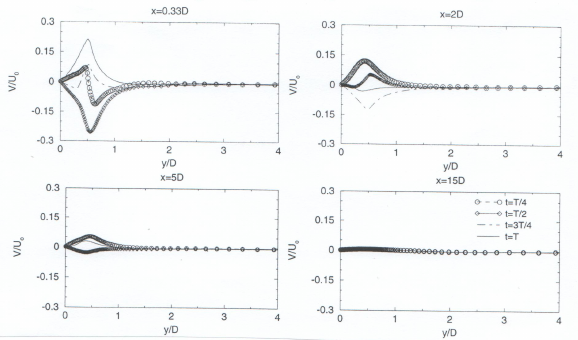}
   \caption{Profiles of the unsteady radial velocity for the transonic jet}
    \end{center}
\end{figure} 

\begin{figure}
  \begin{center}
\includegraphics[width=120mm]{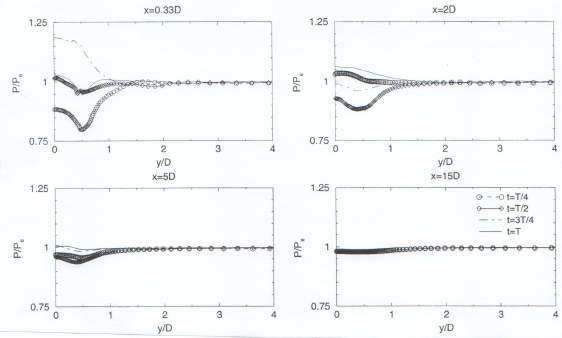}
   \caption{Profiles of the unsteady pressure for the transonic jet}
    \end{center}
\end{figure}

\begin{figure}
  \begin{center}
\includegraphics[width=120mm]{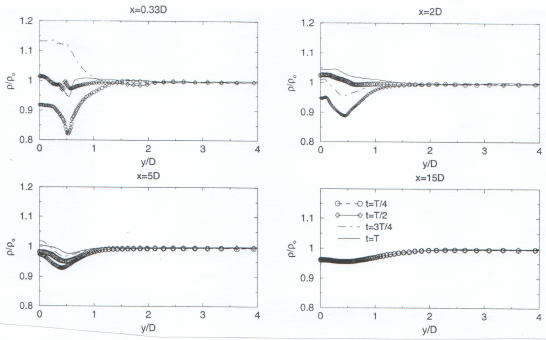}
   \caption{Profiles of the unsteady density for the transonic jet}
    \end{center}
\end{figure} 

\begin{figure}
  \begin{center}
\includegraphics[width=120mm]{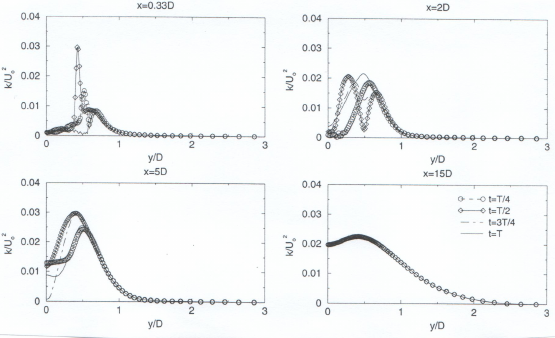}
   \caption{Profiles of the unsteady turbulent kinetic energy for the transonic jet}
    \end{center}
\end{figure} 

\begin{figure}
 \begin{center}
\includegraphics[width=120mm]{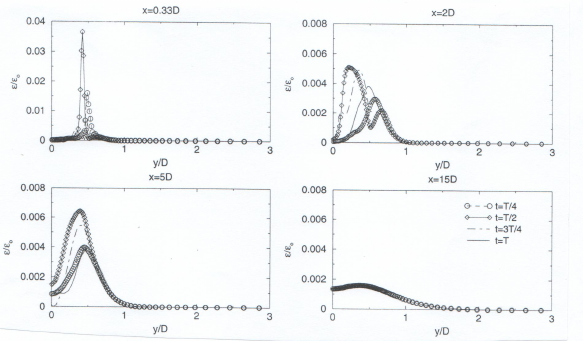}
  \caption{Profiles of the unsteady dissipation rate for the transonic jet (with 
$\epsilon_{o}={U_{o}}^{3}/D$)}
    \end{center}
\end{figure}

\begin{figure}
  \begin{center}
\includegraphics[width=100mm]{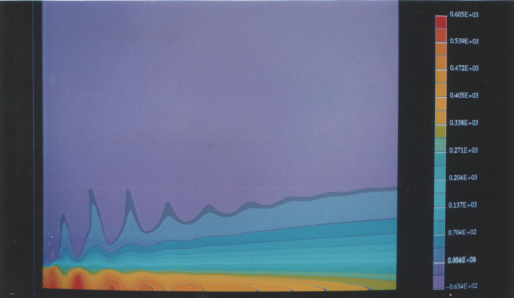}
   \caption{Field of the unsteady streamwise velocity for the supersonic jet}
    \end{center}
\end{figure} 

\begin{figure}
  \begin{center}
\includegraphics[width=100mm]{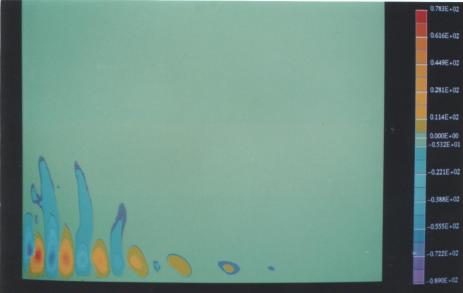}
     \caption{Field of the unsteady radial velocity for the supersonic jet}
    \end{center}
 \end{figure}   

The unsteady profiles of velocity, pressure, density, turbulent energy and dissipation 
rate of turbulent energy are fitted in figures 4 to 9. These 
figures show the unsteady variations for four sections of the mesh located 
at x$=$0.33D, x$=$2D, x$=$5D and x$=$15D, for four moments of a pseudo-period: 
T/4, T/2, 3T/4 and T. A high unsteadiness is active for all quantities in 
the near region at x$=$0.33D and x$=$2D. This flow unsteadiness originates from 
the fluctuations in the shear-layer of the jet. The mixing layer becomes unstable 
near the inlet, due to the Kelvin-Helmholtz instability, then rolls up to form 
vortex rings. This phenomenon has been largely experimentally studied by \cite{LG92} and 
numerically by \cite{VO94} for the incompressible jet. In the present calculations the 
coherent sructures appear very close to the inlet. The numerical results put in 
evidence a strong flow instability for x$=$0.33D and x$=$2D. They are in agreement with the 
experiments of \cite{ASE93} and \cite{LG92} and the numerical study of 
\cite{GOH87} which report the presence of instabilities before two 
diameters. After this location, the vortical structures grow by pairing. \cite{GOH87} 
simulated several pairings in the near-field of a compressible jet. They showed that 
pairings occur at determined locations in the flow, when the resulting vortice is 
too large the potential core is broken up. This process is fundamentally different 
in the round jet compared to the plane jet. As shown in figures 10 and 11, it 
involves the contraction and the expansion of the vortice rings and affects the potential 
core (see also figures 4 to 9) by the induction phenomenon of Biot and Savart. The evolution 
of the coherent structures is largely influenced by this process of contraction and shearing which involves the distorsion of the vortical structures (Fourguette \etal 1991). Moreover, 
this mechanism accelerates the evolution of the secondary three-dimensional structures 
as demonstrated by \cite{LG92}. The present results put in evidence the damping of 
the coherent structures as they are convected downstream (figures 4 to 9). 
The figure 11, where the field of the unsteady radial velocity is visualized, shows clearly 
this decay of the large scale structures. The unsteadiness is strong at x$=$0.33D 
and x$=$2D (see figures 4 to 9), downstream at x$=$5D the coherent structures 
are weakened and at x$=$15D the flow has a steady aspect. The weakening of the 
unsteadiness in the near-field corresponds to the transfer of a subtantial part 
of the coherent instability to random turbulence. So, the decay of coherent 
structures is correlated by the turbulence development. Near the inlet at x$=$0.33D the 
turbulent quantities (figures 8 and 9) are strongly unsteady and the time-averaged 
level (see figure 16) is low. This corresponds to a domination of the coherent 
structures. Downstream at x$=$2D and x$=$5D, the large scale structure dominance 
shades off and the coherent part becomes of the same order then weaker than the 
random component. The damping of the organized unsteadiness is due to the nonlinear 
effects. They are generated 
by the growth of the Kelvin-Helmholtz instability waves, they involve a 
swifter diffusion of the mean flow and the generation of random turbulence 
by the Reynolds stresses. This evolution of the two-dimensional structures is in 
agreement with the experimental investigation of \cite{SKH81}.  These authors 
observed the same features in the evolution of coherent structures in the 
near-field of a round jet. Moreover, they demonstrated that at x/D=1.5 
the structure front is characterized by an intense shearing when the 
drag is a diffusive zone. Downstream, at x/D=4.5, the shearing is weakened by 
the action  of the diffusion mechanisms. In this experiment, up to x/D=4.5 the coherent turbulence 
is gradually diluted but at this location this quantity is larger than the random 
turbulence. According to these authors, at the end of the potential core located at four or 
five diameters the  instability waves reached an adequate magnitude to 
influence the flow. Then the shear-layer and the jet width grow fastly. 
The visualization of the unsteady field of the radial velocity for the 
supersonic jet in figure 11 shows the vanishing of the coherent structures 
before eleven diameters. This result is close to the experimental observations of 
\cite{HZ81} who show that the periodicity of the large scale vortices is lost 
beyond x$=$8D. \cite{LG92} locate this region at x$=$10D where the round jet lost its 
organized aspect. These discrepancies in the exact location of the coherent structure 
vanishing may be explained by different inlet conditions.

\subsection{Flow quantities and interaction with coherent structures}

\subsubsection{Coherent structures and velocity field}

\begin{figure}
  \begin{center}
\includegraphics[width=100mm]{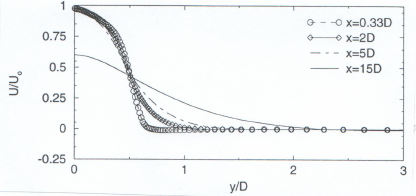}
  \caption{Profiles of time-averaged streamwise velocity for the transonic jet}
    \end{center}
 \end{figure}   

\begin{figure}
  \begin{center}
\includegraphics[width=100mm]{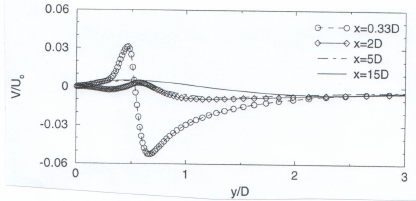}
  \caption{Profiles of time-averaged radial velocity for the transonic jet}
    \end{center}
 \end{figure}

The profiles of streamwise and radial unsteady velocities (figures 4 and 5) show a 
high instability near the inlet at x$=$0.33D and x$=$2D. The visualization 
of the unsteady field of the spanwise velocity in figure 11 puts in evidence an alley 
of vortices moving in the mixing layer already experimentally observed by 
\cite{LF75}. These structures decay  as they move downstream and they disappear in 
the far-field after eleven  diameters. The field of the unsteady streamwise velocity 
(see figure 10) allows the observation 
of expansion and contraction process of the vortice rings. These last ones, located 
in the mixing layer, affect the central zone by the induction phenomenon of 
Biot and Savart. The central region is then characterized by a succession of 
regions of compression and expansion. Downstream, at x$=$5D and x$=$15D (see figures 
4 and 5), the velocity unsteadiness damps and the flow becomes more diffusive which 
involves the velocity decrease and the jet expansion (see figure 12). 

The profiles of unsteady and time-averaged (figures 5 and 13) radial velocities 
put in evidence the high level of the radial velocity in the near region. Its intensity 
decreases downstream at x$=$5D and x$=$15D. The maximum near the inlet proves the 
high entrainment in the near-field. The experimental 
investigation of \cite{HZ81} confirms the strong entrainment in this region. They located 
the maximum of the spanwise velocity between 3.5 and 8 diameters, whereas in our results it 
takes place nearer the inlet. This high entrainment 
has been also observed by \cite{LG92} who established the link with the presence 
of coherent structures. In fact, more the jet is laminar, 
more it is apt to imply a high entrainment. In this flow, the coherent structures 
evolving in the mixing layer influence drastically the entrainment process which 
rises under their effects.

\subsubsection{Interaction with thermodynamic quantities}

\begin{figure}
  \begin{center}
\includegraphics[width=100mm]{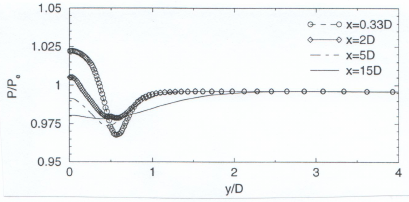}
  \caption{Profiles of the time-averaged pressure for the transonic jet}
    \end{center}
 \end{figure}

A high unsteadiness of the pressure can be seen in the near-field at x$=$0.34D 
and x$=$2D (see figure 6). According to \cite{ZL92} these pressure fluctuations 
are at the source of the instability waves which involve the genesis of the 
coherent structures. The analysis 
of the time-averaged pressure profiles (figure 14) shows a 
minimum in the mixing layer whereas a maximum is located in the central region 
on the axis. This fact has been already reported by \cite{GOH87} in a  
numerical investigation. According to these authors the pressure minima 
coincide with the minima of the normal stress in the radial direction. This points  
out the determinant role of the spanwise velocity fluctuations in the 
generation of the static pressure decrease from the axis until the mixing layer. 
At this location the pressure is generally weaker than the ambiant pressure, 
while a local maximum is located in the potential core where the vortices are 
away. This pressure discrepancy between the shear-layer and the potential core 
is due to velocity gradients. According to \cite{BA67} the variation of the 
time-averaged  pressure compared to the pressure at the equilibrium is function of 
the velocity gradient. 
The succession of minima and maxima of the unsteady pressure in the mixing layer 
is a consequence of the interaction between the pressure field and the coherent 
unsteadiness. The minima correspond to the vortical 
structure centres and the maxima are located between the vortices (Hussain 1986).

The density (see figure 7) presents the same tendancies that the pressure in 
the near-field. A local maximum is situated in the central region while the minimum 
is reached in the shear-layer. The high unsteadiness near the inlet damps 
downstream. When the coherent structures have disappeared after x/D$=$11, the turbulent 
flow is almost fully-developed. Then the diffusion effects rise and the density becomes 
uniform at the level of the surrounding density.

\begin{figure}
  \begin{center}
\includegraphics[width=100mm]{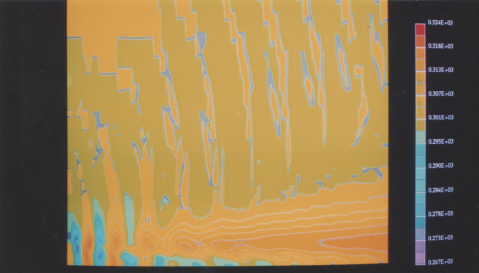}
  \caption{Field of the unsteady temperature for the supersonic jet}
    \end{center}
 \end{figure}  

The field of the unsteady temperature (see figure 15) shows clearly the variations of 
this quantity in the mixing layer at the time of the vortice crossing. The 
temperature increases in the far-field after eleven diameters, this 
corresponds to the progressive degradation of the energy contained in the jet 
by the dissipative mechanisms associated to the mean and turbulent motions.

Finally, a high influence is found in the near-field between the evolution of the 
thermodynamic quantities and the evolution of the organized structures. 
This fact puts in evidence the coupling between these quantities and the coherent 
part of the turbulence.

\subsubsection{Turbulent field}

The analysis of the turbulent field (figures 8 and 9) clues that the shear-layer is 
a region where a high level of turbulence occurs. Indeed, the turbulence is 
generated in this zone where the shearing is intense. As for the other 
quantities (velocity, density and pressure) the variations of 
turbulent energy and of dissipation rate are 
very large in the near-field and disappear progressively downstream. This high 
unsteadiness is associated with the presence of coherent structures. The 
regions located between the large scale vortices are characterized by a strong 
shearing, so the maxima of the turbulent energy occur in these zones. This result 
is in agreement with the theory on coherent structures developed by 
Hussain (1983, 1986). Near the inlet, at x$=$0.33D (figure 8), compared 
to the time-averaged quantities (figure 16), the amplitude of the unsteadiness 
is very high: more than three times the 
time-averaged turbulent energy for the transonic jet. This fact is in agreement with the 
experiments of \cite{HZ81} and \cite{SKH81} where the coherent part of the 
turbulent energy is much greater than the random part in the near-field. This 
indicates a domination of the large scale vortices with respect to the small scale 
turbulence. Between x$=$0.33D and x$=$5D (see figures 8 and 9) the organized unsteadiness 
decays significantly, the coherent structures damp as they are convected downstream. 
The weakening of the large scale vortices is correlated by an increase of the 
time-averaged turbulent energy (see figure 16) between these two locations. The 
growth of the incoherent turbulence is due to the transfer of a subtantial part 
of the organized instability to random turbulence by the 
nonlinear mechanisms of the flow.  After x$=$5D the unsteadiness is weak so the 
turbulence is dominated by the small scales. The residual instability has disappeared 
at eleven diameters. 

On the modelling aspect, the unsteadiness simulated in the near-field  of the compressible 
round jet supports that the equality supposed  between the 
production and the dissipation which is at the source of the classical modelling 
is not verified in the whole jet. If this hypothesis is valid in the fully-developed 
region of a round jet, this is not the case in the near-field as 
shown by \cite{RO72} in an experimental investigation. The presence of coherent 
structures alters strongly the near-field. The energy spectra is not equilibriated 
in this region due to the presence of peaks (e.g. Ha Minh \& Kourta, 1993) which correspond 
to the coherent structures.

\begin{figure}
  \begin{center}
\includegraphics[width=100mm]{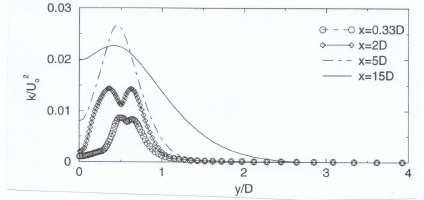}
  \caption{Profiles of the time-averaged turbulent energy for the transonic jet}
    \end{center}
 \end{figure}  

The mean level of the turbulent kinetic energy is low in the near-field 
(figure 16) especially in the central region. In 
this zone the velocity gradients are weak, a low production is involved which explains 
the central depression predicted in the turbulent energy profiles. The turbulent energy 
is produced in the mixing layer where a high shearing occurs between the vortices. 
This agrees with the experimental results of \cite{PL93} where the production 
is in majority some shear production. The presence of a central depression has also 
been observed by \cite{BP79}. In the present 
simulations the depression is progressively filled 
downstream by the extension of the mixing layer and the resulting 
diffusion. The turbulent 
energy maximum has been located in the 
mixing layer between 2.5D and 7D by \cite{HZ81} and at about 6D by \cite{BP79}. In the 
present study the maximum occurs at 9.5D for the subsonic flow (Mach 0.3). In fact, 
the location of the maximum depends on both inlet conditions and Reynolds 
number according to \cite{BP79}.

\subsection{Compressibility effects}

\subsubsection{On preferred mode}

\begin{figure}
  \begin{center}
\includegraphics[width=100mm]{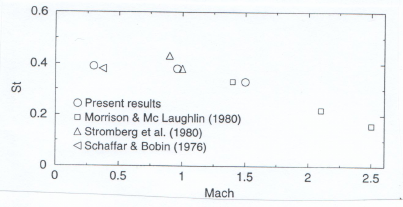}
  \caption{Evolution of the Strouhal number with the Mach number}
    \end{center}
 \end{figure}

Flow computations for Mach numbers of $0.3$, $0.96$ (see figures 4 to 9) 
and $1.5$ show the near-field domination by the organized unsteadiness 
(e.g. Reynier 1995). These instabilities are quasi-periodic (see figure 2) 
and square with the preferred mode found by \cite{CC71}. The present results 
on near-field vortices allow the evaluation of the Mach number influence on the 
Strouhal number associated to the dominant instability. 
The figure 17 represents the Strouhal numbers measured by \cite{SB76}, Stromberg, McLaughlin 
\& Troutt (1980) and \cite{MM80} and the present computation results. There is 
an excellent agreement between our numerical predictions and the values provide by 
experiments for the transonic and supersonic jets. The value predicted for the 
quasi-incompressible jet is very close to that of 0.38 measured by \cite{SB76} for an exit 
velocity of 130m/s. For the simulations with the Mach numbers of 0.3 and 
0.96 the Strouhal number stays quasi-firm. For this velocity range, 
the compressibility effects remain weak and the turbulence structure is not fundamentally 
altered. It is not the same for the supersonic jet where the convective Mach number is 
larger. In our calculations the Strouhal number of the dominant component decays when the 
Mach number increases (its value is 0.38 for Mach 0.96 and 0.33 for Mach 1.5). This agrees 
with the experimental investigations of \cite{LUBB87} and \cite{MM80}. The Strouhal 
number decrease with the Mach number growth is to put together with the theory of 
\cite{PR88} on compressible mixing layers. The decrease of the Strouhal number
corresponds to a diminution of the number of coherent structures evolving in the jet. This places 
in a prominent position the stabilizing effect of the Mach number on the flow. 
The diminution of the dominant mode of the 
Strouhal number for the compressible flows is caused, according to \cite{MI58}, 
by a decrease of the growth rate of the Kelvin-Helmholtz instability.

\subsubsection{On turbulence and coherent structures}

\begin{table}
  \begin{center}
       \begin{tabular}{cccc}
Mach number & Time-variation $\Delta \overline{\rho} \over \rho_{0}$  
& Time-variation $\Delta \tilde{k} \over {U_{0}}^{2}$ & 
Time-variation  $\Delta \tilde{V} \over U_{0}$ \\[3pt] 
$0.3$ & $0.018$ & $0.016$ & $0.35$ \\
$0.96$ & $0.18$ & $0.02$ & $0.317$ \\
$1.5$ & $0.38$ & $0.0195$ & $0.256$ \\
     \end{tabular}
     \caption{Maximum time-variations of several flow quantities in the mixing layer at 
x$=$1.5D and y$=$0.53D}
  \end{center}
\end{table}

In table 2, the maximum adimensional time-fluctuations of density, turbulent energy and radial 
velocity are compared for the three computed cases at x$=$1.5D and y$=$0.53D. 
The table shows a growth of the time-fluctuations for the density with the Mach number. 
This seems logical because the compressibility effects are more important for the transonic 
and supersonic jets. The adimensional time-variations decrease for the radial velocity with 
the Mach number but the absolute variations rise. For the turbulent kinetic 
energy the adimensional time-fluctuation is quasi-firm. This shows a growth of the 
absolute fluctuations with the square of the Mach number. This indicates a dependence of the 
coherent structure characteristics on the Mach number. \cite{ASE93} also observed 
the increase in strengh of the large scale vortices with the Mach number. In their experimental 
investigation on a compressible round jet \cite{HZ81} established that the large 
scale structures depend on the Reynolds number. By splitting the Reynolds 
stresses in two components, \cite{HZ81} pointed out that the coherent part of the 
turbulence becomes larger for high Reynolds numbers. They also remarked that 
the coherent structures become more energetic. The momentum 
transport across the shear-layer is then larger: this agrees with the 
increase of turbulent energy and radial velocity time-variations for the 
high Mach numbers in the present numerical simulations.

\begin{figure}
  \begin{center}
\includegraphics[width=100mm]{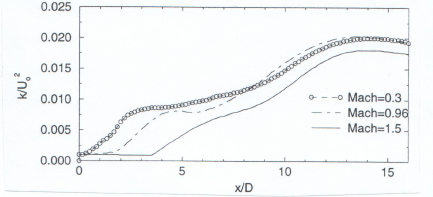}
 \caption{Axis evolution of the time-averaged turbulent energy for the three 
computed cases}
    \end{center}
 \end{figure}

The evolution of the time-averaged turbulent energy on the axis for the 
different simulations 
is plotted in figure 18. The axis profiles of the turbulent energy level off 
in the near-field which corresponds to the potential core region. The plateau 
where the turbulent energy is low is due to a weak production in the central 
region of the jet. The length of the plateau rises with the Mach number. \cite{MM80} 
observed in their experiment on supersonic jets the dependence of the potential core 
length on both Mach and Reynolds numbers. Its length increases with the first and 
decreases with the second. The first point is well correlated by our numerical results. 
Downstream the region where the turbulent energy levels off, the curves show a 
strong growth of the turbulent energy. This is a consequence of the transfer of a 
coherent turbulence subtantial part to random turbulence. This shows, in 
agreement with \cite{SKH81}, that the end of the potential core is characterized by 
a drastic change in the turbulence structure. The increase of the turbulent energy 
originates from the growing of the mixing layer which involves 
the diffusion of the turbulence from this region to the whole 
jet. After 10D the turbulent energy (see figure 18) follows the same evolution 
for the three computed cases, the compressibility effects are 
attenuated in this region of the flow where the coherent structures have vanished 
(see \S\,4.1.2).

\subsubsection{Influence on expansion rate}

\begin{figure}
  \begin{center}
\includegraphics[width=100mm]{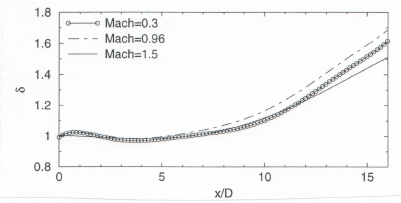}
  \caption{Evolution of the expansion rate with the Mach number}
    \end{center}
 \end{figure}

In figure 19 the expansion rates computed for the three Mach numbers are 
represented. If the Mach number 
has a stabilizing effect on the flow, the expansion rate should decrease 
when the Mach number increases. Paradoxically, the jet has a greater expansion 
at Mach $0.96$ than at Mach $0.3$, whereas the expansion rate is lower for the 
supersonic case. The compressibility effects are 
weak for the low values of the convective Mach, 
then the growth rate of the Kelvin-Helmholtz instability has not been 
largely reduced between Mach numbers of $0.3$ and $0.96$. But the Reynolds number has 
slightly increased which explains the expansion rate growth. This phenomenon is 
directly linked to the characteristics of the coherent structures. According to 
\cite{HZ81} they depend on the Reynolds number. The momentum 
transport across the shear-layer is greater at high Reynolds numbers: this implies a 
rise of diffusion and expansion rate. In the present 
predictions the Strouhal number is quasi-firm between the Mach 
numbers of $0.3$ and $0.96$. This indicates that the same number of coherent 
structures evolves in the near-field. But, the Reynolds number has increased  
so the diffusion is more important which explains the higher expansion rate for 
the jet at Mach $0.96$.

In agreement with the survey of the jet instability theory by \cite{MI84}, 
the expansion rate is weaker for the supersonic jet. This is a consequence of 
the Strouhal number diminution (see figure 17). Indeed, the presence of coherent 
structures has for effect an increase of the flow interface, when the Strouhal 
number reduces, the interface length becomes shorter and the diffusion in the flow 
is weaker. This point explains the reduction of the expansion rate for the supersonic 
jet in our simulations. As in the paragraph \S\,4.3.1 where the Mach number effect 
implies a Strouhal number decrease, the diminution of the expansion rate for 
the supersonic jet shows a stabilization of the jet by the compressibility 
effects. This agrees as previously with the theory of \cite{PR88}. 
The lessening of the expansion rate for the supersonic Mach numbers has been 
already experimentally 
investigated in coaxial jets by Schadow, Gutmark \& Wilson (1989) who have 
established the dependence of the expansion rate on the convective Mach number.

\section{Conclusion}

The results presented in this paper have been obtained for compressible 
jets with an under-developed turbulence at the inlet. In this compressible 
flow, the use of the semi-deterministic modelling allows the simulation of the 
unsteadiness corresponding to the coherent structures which evolve in the near-field 
and dominate the mixing layer of the jet. The agreement between the Strouhal numbers 
of the preferred mode corresponding to the dominant instability predicted by the 
calculations for several Mach numbers and those provided by previous 
experiments is excellent. Moreover, the evolution of the large scale vortices 
and their interaction with the unsteady fields of velocity, pressure and turbulent 
quantities predicted by the simulations show the same features than in previous 
experiments. The instability appears very close to the inlet, the unsteadiness 
is strong at x$=$0.33D, and decays by moving downstream. The coherent structures 
vanish at eleven diameters which agrees with other experimental investigations. 
The profiles of the time-averaged pressure show its decrease from the axis until 
the mixing layer. The pressure variations in the mixing 
layer at the time of the coherent structure crossing have been also simulated. 
These results agree with previous numerical simulations and theory on coherent 
structures. The unsteadiness of the turbulent quantities 
shows a strong coupling with the evolution of the large scale vortices. The coherent 
turbulence dominating the near-field is transfered to random 
turbulence under the action of the nonlinear mechanisms of the flow. In 
consequence, the growth of the random turbulence is correlated by the damping of 
the coherent structures.

The results on the influence of compressibility effects on dominant instability 
and expansion rate are in agreement with previous theory on the stabilization of flows 
by the Mach number increase. The Strouhal number of the preferred mode reduces 
with the Mach number growth. The expansion rate depends on the coherent structures, 
it rises with the Reynolds number and decreases with the Mach number for the supersonic 
jet. The large scale vortices become more energetic for high Mach numbers but they 
are more rarely which explains the decrease of the expansion rate.

If this study is proving conclusive, some improvements of the method applied 
here are possible. The prediction of the location of the turbulent energy maximum in 
the jet is not in fair agreement with experiments. If the inlet conditions are not the same 
this cannot absolutly justify this discrepancy. In the model used, the constants have 
been recalibrated on a backward-facing step and not on a jet. To escape of a new 
calibration for every flow, a solution is to use functions at the place of 
constants to take account of the organized unsteadiness. This involves a turbulence 
model more nonlinear, therefore a numerical scheme more unstable. Another solution 
would be the use of a model more elaborated as a Reynolds stress model to 
improve the result precision. Therefore, the method applied in this paper allows the best hopes to the simulation 
of coherent structures for flows at high Reynolds numbers and constitutes an 
alternative to the LES.

\hspace{0.5in}

Financial support for this study was provided by the Soci\'et\'e Europ\'eenne 
de Propulsion  under contract PRC/CNRS n$^{o}$90.0018C. We are also grateful 
to the CNES which supports this work under the post doctorate grant of P. Reynier.

\end{document}